\documentclass[aps,twocolumn,showpacs,floatfix]{revtex4} 
\usepackage{amsmath,amssymb,graphicx,times}

\newcommand{\be}{\begin{equation}}
\newcommand{\ee}{\end{equation}}
\newcommand{\bea}{\begin{eqnarray}}
\newcommand{\eea}{\end{eqnarray}}
\newcommand{\pavg}{\Pi_M(t)}

\begin{document}


\title{Coherent Exciton Dynamics and Trapping in Topologically Disordered
Systems
}
\author{Oliver M{\"u}lken and Alexander Blumen}
\affiliation{
Theoretische Polymerphysik, Universit\"at Freiburg,
Hermann-Herder-Stra{\ss}e 3, 79104 Freiburg, Germany}

\date{\today} 
\begin{abstract}
We analyze the coherent dynamics of excitons in three dimensional
topologically disordered networks with traps. If the interactions between the
nodes of the network are long ranged, i.e., algebraically decaying as a
function of the distance between the nodes, the average survival
probability of an exciton surprisingly shows a characteristic decay with
features similar to the decay found for regular one-dimensional systems.
We further show how this decay can be related to the eigenstates of the
same system without a trap.
\end{abstract}


\maketitle


\section{Introduction}

Coherent dynamical processes are of interest in many fields of research.
Naturally, quantum mechanical systems like atoms in optical lattices
display coherent dynamics, see \cite{bloch2005} and references therein.
Optical systems, for instance, lattices of evanescently coupled waveguides
allow to study the basic principles of coherent energy transfer from the
experimental side, given that these systems have long decoherence times
\cite{hagai2008}. Also some biological processes show
instances of being coherent: an example is the highly efficient energy
transfer in photosynthesis, which cannot be understood based only on
(incoherent) diffusion \cite{fleming2004,cheng2006,engel2007}. 

The transport takes place over many, quite diverse geometries, ranging
from very ordered lattices to random networks of active sites.  Here, an
efficient approximate description of the dynamics is given by a
tight-binding ansatz which also allows to take the structure over which
the excitation dynamics occurs fully into account
\cite{mb2006b,mbb2006a,mpb2007a}. Depending on the system, the
interactions between sites extend from nearest-neighbor couplings to being
of long-range type. An example for systems which allow to experimentally control and
manipulate the range of the interactions are ultra-cold Rydberg gases
\cite{anderson1998,mourachko1998}. Furthermore, it is possible to excite
individual atoms locally; at later times these excitations
may spread over the whole aggregate.  By considering only the original
Rydberg state and the state the local excitation is in, one can view the
whole gas of atoms as being a collection of two-level systems
\cite{westermann2006}; this resembles the situation in so-called
spin-gases \cite{calsamiglia2005}.  The typical interactions between the
atoms of the Rydberg gas is of dipole-dipole nature, decaying with the
third power of the mutual distance between the atoms.

Now, instead of directly following the propagation of a single excitation
one can monitor the (global) transport process. One can introduce
dissipation into the system in a controlled way. So-called excitation
traps, which are placed at certain sites of the network, will
``absorb''/``trap'' the excitation once it hits the trap.  For Rydberg
atoms such traps can be realized by exciting one atom into a different
state such that the excitation gets to be absorbed at this site
\cite{mbagrw2007,reetz-lamour2008}. By monitoring the decay of the
survival probability for the excitation to remain inside the system, i.e.,
not to get absorbed by a trap, one obtains information about the
dynamics within the system. 

The paper is organized as follows: After briefly recalling the
phenomenological approach to excitation trapping, we show how the exciton
survival probability for a given system can be related to the eigenstates
of the same system without traps. Our model system, a random assembly of
two-level systems in a cubic box, is described next, before presenting our
numerical results, which were obtained using LAPACK routines.

\section{Coherent exciton trapping}

The Hamiltonian for a system of $N$ nodes without traps is ${\bf H}_0$,
where the states $|j\rangle$ are associated with excitations localized at the
nodes $j$ ($j=1,\dots,N$). These states are assumed to form a complete, orthonormal
basis set of the whole accessible Hilbert space ($\langle k | j \rangle =
\delta_{kj}$ and $\sum_k |k~\rangle\langle~k| = {\bf 1}$).  Within a
phenomenological approach, the Hamiltonian, which incorporates trapping of
excitations at the nodes $m$ ($m\in{\mathcal M}$ and ${\mathcal
M}\subset\{1,\dots,N\}$), is given by ${\bf H} \equiv {\bf H}_0 - i{\bf \Gamma}$,
where $i{\bf \Gamma} \equiv i \Gamma \sum_{m\in{\mathcal M}} | m \rangle
\langle m |$ is the trapping operator, see Ref.~\cite{mbagrw2007} for
details. As a result, ${\bf H}$ is non-hermitian and has $N$ complex
eigenvalues, $E_l = \epsilon_l - i\gamma_l$ ($l=1,\dots,N$) where
$\gamma_l>0$, and $N$ left and $N$ right eigenstates, denoted by
$|\Psi_l\rangle$ and $\langle\tilde\Psi_l|$, respectively.  The transition
probabilities are given by
\be
\pi_{kj}(t) = \big|\sum_l \exp(-\gamma_lt) \exp(-i\epsilon_lt)\langle k |
\Psi_l \rangle \langle \tilde\Psi_l | j \rangle\big|^2,
\ee
so that the imaginary parts $\gamma_l$ of $E_l$ determine the temporal
decay. 

The mean survival probability $\pavg$ of an excitation in the presence of
$M$ trap nodes is a global property of the network and is defined as
\be
\pavg \equiv \frac{1}{N-M} \sum_{j\not\in{\mathcal M}} \sum_{k\not\in{\mathcal M}}
\pi_{kj}(t),
\ee
i.e., $\pavg$ is the average of $\pi_{kj}(t)$ over all initial nodes $j$
and all final nodes $k$, neither of them being a trap node. For
intermediate and long times and for a small number of trap nodes, $\pavg$
is mainly a sum of exponentially decaying terms \cite{mbagrw2007}:
\be
\pavg \approx \frac{1}{N-M} \sum_{l=1}^N \exp(-2\gamma_lt).
\label{piapprox}
\ee 
If the imaginary parts $\gamma_l$ obey a power-law with an exponent
$1/\eta$ ($\gamma_l \sim a l^{1/\eta}$), the mean survival probability
scales at intermediate times as $\pavg \sim t^{-\eta}$.

In the subsequent calculations we will use the following (approximate)
average survival probability for a single trap ($M=1$), obtained by
slightly modifying Eq.~(\ref{piapprox}), 
\be
\Pi(t) \equiv  \frac{1}{N} \sum_{l=1}^N \exp(-2\gamma_lt),  
\label{pit}
\ee
which ensures the same normalization for different $N$, i.e., $\Pi(0) = 1$
for all $N$.

\subsection{Relation to the eigenstates of ${\bf H}_0$}\label{trap-eigenstate}

In a perturbation theoretical treatment, assuming the trapping strength
$\Gamma$ to be small, the imaginary parts $\gamma_l$ can be related to the
eigenstates $|\Psi_l^{(0)}\rangle$ of the system without traps
\cite{mpb2008b}. For a single trap at node $k$ one has 
\be
\gamma_l = \Gamma
\big|\langle k| \Psi_l^{(0)}\rangle \big|^2.
\ee 

In the continuum limit the index $l$ becomes a continuous variable, say, $x$ and
the imaginary parts lie in the interval $\gamma_l \in [0, \gamma_{\rm
max}]$. If we know the density $\rho[\gamma(x)] = dx(\gamma) / d\gamma$ of
the $\gamma(x)$, Eq.~(\ref{pit}) can be recast into the form
\be
\Pi(t) = \int\limits_0^\infty d\gamma \ \rho(\gamma) \ \exp(-2\gamma t)
\equiv {\mathcal L}\{\rho(\gamma)\},
\label{pi-cont}
\ee
where ${\mathcal L}\{\rho(\gamma)\}$ denotes the Laplace transform of
$\rho(\gamma)$.  On the other hand, if we know $\Pi(t)$, we can obtain the
density $\rho(\gamma)$ by the inverse Laplace transform ${\mathcal
L}^{-1}\{\Pi(t)\}$. 

Take as an example a linear system with a trap at one end. At intermediate
times, the mean survival probability will decay as $\Pi(t) \sim t^{-1/2}$
\cite{mpb2008b}. From Eq.~(\ref{pi-cont}) one immediately obtains
$\rho(\gamma) \sim \gamma^{-1/2}$ and from this by integration $\gamma
\sim x^{2}$. We will elaborate on the relation between trapping and the
eigenstates of a closed system elsewhere in more detail \cite{mb2008b}.

\section{Quantum network}

\begin{figure}[htb]
\includegraphics[clip=,width=\columnwidth]{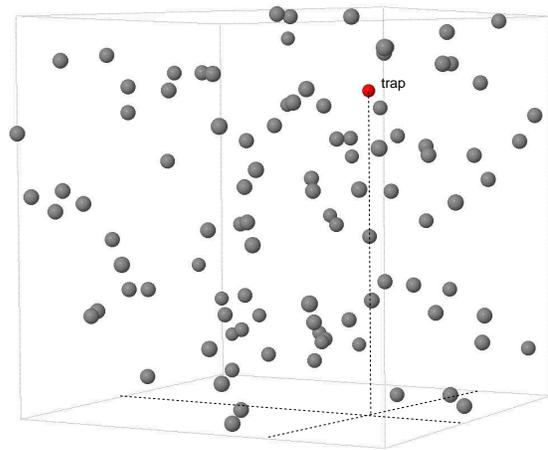}
\caption{(Color online) Exemplary random configurations of $N=100$ nodes in a cube, one
of which is a
trap (labeled red dot). 
}
\label{configuration_n0100}
\end{figure}

We start from a random configuration of $(N-1)$ identical nodes and {\sl
one} trap node.  All $N$ nodes are placed at random in a $3$-dimensional
box with Cartesian coordinates $\{x^{(i)}\}$, with $i=1,2,3$. Then the
distance between two nodes $j$ and $k$ is given by 
\be
\Delta_{j,k} = \Bigg[\sum_{i=1}^3
\big(x_j^{(i)}-x_k^{(i)}\big)^2\Bigg]^{1/2},
\ee
where the coordinates $x_j^{(i)}$ and $x_k^{(i)}$ are homogeneously
distributed random numbers in the interval $[0,N]$. To relate our results to
the energy transfer dynamics within Rydberg gases we consider interactions
decaying as $\Delta_{j,k}^{-3}$. In the absense of traps the corresponding
Hamiltonian ${\bf H}_0$ has the following matrix elements
\be
\langle k | {\bf H}_0 | j\rangle = 
\begin{cases}
\displaystyle -\Delta_{j,k}^{-3} & \mbox{for} \ k\neq j \\
\displaystyle \sum_{k\neq j} \Delta_{j,k}^{-3} & \mbox{for} \ k=j.
\end{cases}
\label{hamil_long}
\ee
We now choose one of the $N$
nodes to be a trap, i.e., for this node the full Hamiltonian ${\bf H}$ has
an additional purely imaginary matrix element $-i\Gamma$. Since the
configuration of nodes is random, we can (without any loss of generality)
assume in the following that the node labeled $1$ is the trap.
Figure~\ref{configuration_n0100} shows one realization of the system in
which the trap node is explicitly labeled.

For disordered systems, we calculate averages over $R$ different
realizations following 
\be
\big\langle \cdot \big\rangle_R \equiv \frac{1}{R} \sum_{r=1}^R \big[\cdot
\big]_r,
\ee 
where $\big[ \cdot \big]_r$ denotes the realization $r$. In our
calculations we also assume the trapping strength $\big[\Gamma\big]_r$ to
be realization dependent, because we require it to be proportional to the
diagonal element $\big[\langle 1 | {\bf H}_0 | 1 \rangle\big]_r$ in that particular
realization, namely $\big[\Gamma\big]_r \equiv \Gamma \big[\langle 1
| {\bf H}_0 | 1 \rangle\big]_r$. As it will turn out, the dependence of
the decay on the value of $\Gamma$ is quite weak - different $\Gamma$
mainly rescale the time axis.  Thus we will consider only the two extreme
cases: (a) $\Gamma=10^{-6}$, for which a perturbation theoretical
treatment can be justified, and (b) $\Gamma=1$, such that the average
trapping strength is of the same order as the diagonal elements of ${\bf
H}_0$ at the node of the trap.

While each realization $\big[\Pi(t)\big]_r$ corresponds to a specific
spectrum of the $\big[\gamma_l\big]_r$, the correspondence of $\langle
\Pi(t) \rangle_R$ to the average $\langle\gamma_l\rangle_R$ is not that
straightforward.  However, for all $t$, the function $\exp(-2\gamma_l t)$
is convex, therefore, Jensen's inequality applies, see paragraph 12.41 of
\cite{gradshteyn}, such that we obtain for a given $l$
\be
\big\langle \exp(-2\gamma_l t) \big\rangle_R \geq \exp\big(-2 t
\langle\gamma_l\rangle_R \big).
\ee
From this we get a lower bound for $\big\langle \Pi(t) \big\rangle_R$:
\be
\big\langle \Pi(t) \big\rangle_R \geq \frac{1}{N} \sum_{l=1}^N \exp\big(-2
t \langle\gamma_l\rangle_R \big).
\label{pi_lb}
\ee

\section{Numerical results}

For a linear system with equal distance between neighboring nodes it has been
shown in Refs.~\cite{mbagrw2007,mpb2008b}, that the decay of $\pavg$ shows
characteristic features of quantum transport, such as an algebraic decay
of the survival probability at intermediate times. It is not a priori clear
that these features will persist in higher dimensions or in disordered
systems with long-range interactions. As we proceed to show, the
quantum network for a disorded system in three dimension, as defined in
the previous section, still displays, in the ensemble average, distinct
temporal regions with different behaviors of the decay of $\langle \Pi(t)
\rangle_R$.

\begin{figure}[htb]
\includegraphics[clip=,width=\columnwidth]{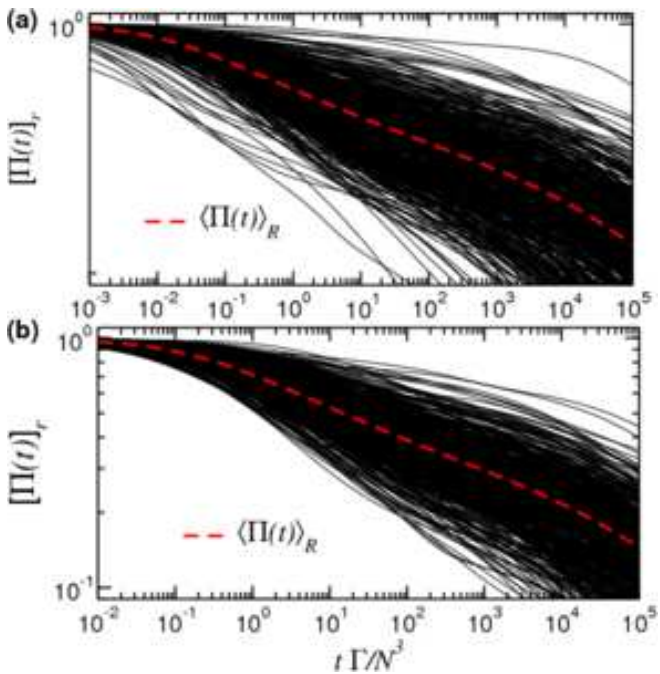}
\caption{(Color online) Decay of $\big[\Pi(t)\big]_r$ as a function of
rescaled time $t\Gamma/N^3$ for $R=500$ different realization and $N=100$:
(a) for $\Gamma=10^{-6}$ and (b) for $\Gamma=1$. The dashed red curves
show the ensemble averages $\langle \Pi(t) \rangle_R$.}
\label{probs_n100}
\end{figure}

As an example, Fig.~\ref{probs_n100} shows $\big[\Pi(t)\big]_r$ obtained
from Eq.~(\ref{pit}) for $N=100$ and for $R=500$ different realizations.
Fig.~\ref{probs_n100}(a) gives the results for $\Gamma=10^{-6}$ and
Fig.~\ref{probs_n100}(b) for $\Gamma=1$. In both figures the average
$\langle \Pi(t) \rangle_R$ is shown as a dashed red line. Obviously, the
various realizations differ strongly in their behavior, displaying
stronger and weaker decays. However, in the ensemble average, the decay of
$\langle \Pi(t) \rangle_R$ is qualitatively comparable to the one found
for regular (one-dimensional) systems, in that it follows a power-law at
intermediate times. Additionally, all realizations show monotonic decays,
although with variations over several orders of magnitude in time.

\begin{figure}[htb]
\includegraphics[clip=,width=\columnwidth]{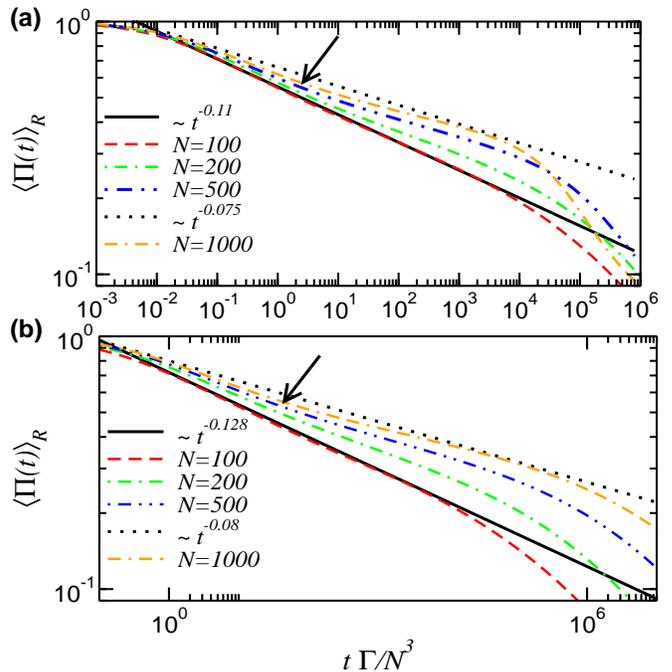}
\caption{(Color online) Ensemble averages $\langle \Pi(t) \rangle_R$ for
different $N$: (a) for $\Gamma=10^{-6}$ and (b) for
$\Gamma=1$. The scalings in the intermediate time regions for $N=100$ and
$N=1000$ are shown as solid black lines along with the appropriate scaling
law. The arrows are guides to the eye pointing at the bend of
$\langle \Pi(t) \rangle_R$ for $N=1000$.
}
\label{probs_avg_n}
\end{figure}

An analysis of $\langle \Pi(t) \rangle_R$ is shown in Fig.~\ref{probs_avg_n}
at the two values $\Gamma=10^{-6}$ and $\Gamma=1$ for different $N$. In all
cases the
intermediate time decay can be fitted by a power-law 
\be
\langle \Pi(t)
\rangle_R \sim t^{-\eta(N)},
\ee 
where, different from the regular linear case \cite{mbagrw2007,mpb2008b},
the exponent $\eta$ is now $N$-dependent. It turns out that approximating
the exponent by 
\be
\eta(N) = \eta_0 N^\mu,
\label{etaN}
\ee
while keeping $\Gamma$ fixed,
yields satisfactory results. Note, however, that $\eta_0$ and also $\mu$
might still depend on $\Gamma$. 

Now, we can estimate $\mu$ based on the results of $\eta$ for $N_1=100$
and $N_2=1000$. From $\eta_0 = \eta(N_1)/N_1^\mu = \eta(N_2)/N_2^\mu$ one
obtains
\be
\mu = \frac{\ln\eta(N_2) - \ln\eta(N_1)}{\ln N_2 - \ln N_1}.
\ee
From the numerical values given in Fig.~\ref{probs_avg_n} we get
approximately $\mu\approx-0.166$ and $\eta_0=0.0349$ for $\Gamma=10^{-6}$
and $\mu\approx-0.204$ and $\eta_0=0.0313$ for $\Gamma=1$. 

From Fig.~\ref{probs_avg_n} as well as from Eq.~(\ref{etaN}) we see that
the exponent $\eta(N)$ decreases with increasing $N$.  Certainly, if
$N$ becomes very large it is quite improbable (in the ensemble average)
for an exciton to encounter the single trap. Therefore, the decay of
$\langle \Pi(t) \rangle_R$ can only be observed at very long
times.

\subsection{Relation to the eigenstates $|\Psi_l^{(0)}\rangle$}

The fact that there is a decay at all can be explained by recalling that
for small $\Gamma$, the decay rates $\gamma_l$ are related to the
eigenstates $|\Psi_l^{(0)}\rangle$ of the system without traps, see above,
which in the ensemble average we can write 
\be
\langle\gamma_l\rangle_R = \left\langle \Gamma |\langle k | \Psi_l^{(0)}
\rangle |^2 \right\rangle_R.
\ee
Now, imagine there are states $|\Psi_l^{(0)}\rangle$ which have no overlap
with the trap node at position $k$. This would cause vanishing $\gamma_l$
for some $l$, which would lead to a non-vanishing $\Pi(t)$ even in the
asymptotic time limit, see Eq.~(\ref{pit}). From Fig.~\ref{probs_n100}
one sees that at least in the displayed time interval all
$\big[\Pi(t)\big]_r$ decay strictly monotonically. Therefore, there are several
eigenstates $|\big[\Psi_l^{(0)}\big]_r\rangle$ which have non-vanishing
overlap with the trap node.

\begin{figure}[htb]
\includegraphics[clip=,width=\columnwidth]{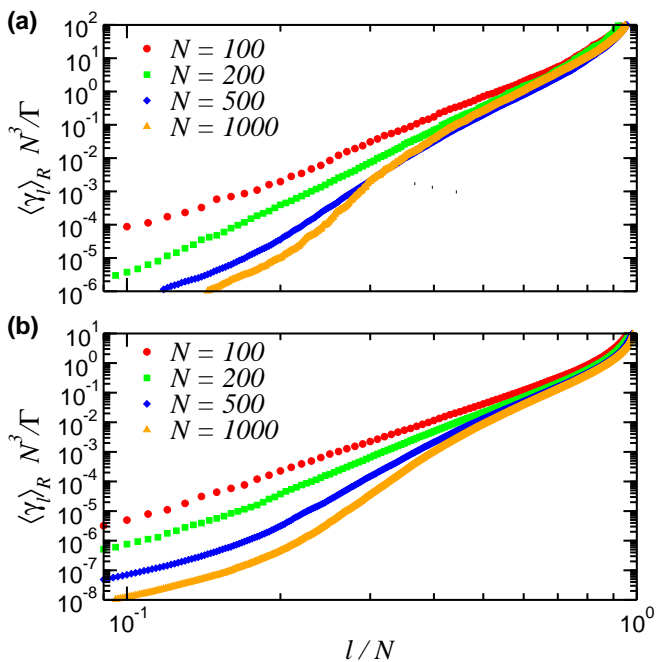}
\caption{(Color online) Ensemble average of the imaginary parts
$\langle\gamma_l\rangle_R$ for different $N$ sorted in ascending order:
(a) for $\Gamma=10^{-6}$ and (b) for $\Gamma=1$.  }
\label{evals_avg_n}
\end{figure}

In Fig.~\ref{evals_avg_n} the average $\langle\gamma_l\rangle_R$ is shown
as a function of $l/N$ (on double-logarithmic scales) for different $N$
and (a) for $\Gamma=10^{-6}$ and (b) for $\Gamma=1$.  One sees that the
$\langle\gamma_l\rangle_R$ have a much richer structure than for the
linear system, with evenly spaced nodes (compare to Fig.~1 of
Ref.~\cite{mpb2008b}).  While for small $N$ a single power-law fit for
intermediate values of $l/N$ can be justified (see curves for $N=100$ and
$N=200$), this ceases to be the case for larger $N$. In fact, one can
identify different regions where different power-laws might hold, see
curves for $N=500$ and $N=1000$.

Due to the fact that Eq.~(\ref{pi_lb}) is only a lower bound, relating
exponents from power-law fits to $\langle\gamma_l\rangle_R$ to the ones
obtained for $\langle \Pi(t) \rangle_R$ [see Fig.~\ref{probs_avg_n}] is
not very conclusive. 
Jensen's inequality [Eq.~(\ref{pi_lb})] holds for the relation
between $\langle \Pi(t) \rangle_R$ and $\langle\gamma_l\rangle_R$ and thus
only allows for qualitative statements. However, it is possible to obtain
quantitative relations between $\langle \Pi(t) \rangle_R$ and the average
density of eigenvalues $\langle \rho(\gamma_l)\rangle_R$. In the continuum
limit we get $\langle \rho(\gamma)\rangle_R = {\mathcal L}^{-1} \{\langle
\Pi(t) \rangle_R\}$. Given a scaling of $\langle \Pi(t) \rangle_R \sim
t^{\eta(N)}$ one has
\be
\langle\rho(\gamma)\rangle_R \sim
\gamma^{\eta(N)-1}. 
\ee

\begin{figure}[htb]
\includegraphics[clip=,width=\columnwidth]{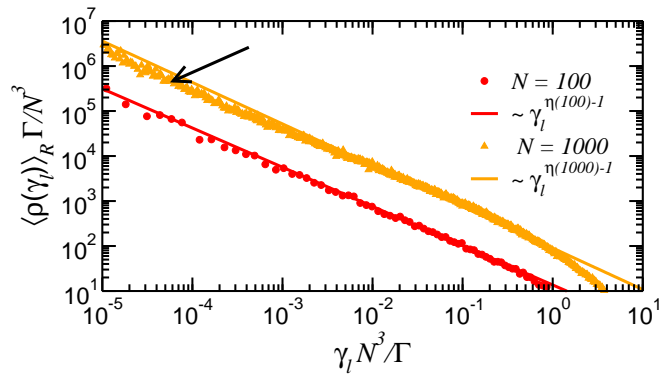}
\caption{(Color online) Density of $\langle\gamma_l\rangle_R$ for $N=100$
(red circles) and $N=1000$ (orange triangles) and for $\Gamma=1$. The
solid lines show the scaling law which follows from the inverse Laplace
transform ${\mathcal L}^{-1} \{\Pi(t)\}$ with $\Pi(t) \sim t^{-\eta(N)}$.
The arrow is a guide to the eye pointing at the bend in $\rho(\gamma)$ for
$N=1000$.  }
\label{densevals_avg_n_g1}
\end{figure}

Figure~\ref{densevals_avg_n_g1} shows the average density
$\langle\rho(\gamma_l)\rangle_R$ (obtained by inversion of $\langle
\gamma_l \rangle_R$) for $N=1000$ and $\Gamma=1$ along with a power-law
fit for intermediate values of $\gamma$. Indeed, the exponent of the
power-law fit can be obtained from $\langle\Pi(t)\rangle_R$. As it turns
out, the best fit is given by $\langle\rho(\gamma_l)\rangle_R \sim
\gamma_l^{\eta(N)-1}$, where $\eta(N)$ is given above for the scaling of
$\langle\Pi(t)\rangle_R$ in the intermediate time region.  Moreover, also
the slight bending of $\Pi(t)$ for $N=1000$ (see arrows in
Fig.~\ref{probs_avg_n}) is recovered in the
corresponding density $\rho(\gamma)$.

\section{Conclusions}

We have calculated the decay of the average survival probability for
coherent excitons in three dimensional topologically disordered systems
with long-range interactions. As for a regular one dimensional system, in
the ensemble average there is a characteristic decay at intermediate
times. We established a lower bound to the decay by employing Jensen's
inequality and further related the average decay rates to the average
eigenstates of the same system without traps. Especially this last point
might be exploited for (indirect) measurements of the eigenstates: For
instance it is possible to control dissipation due to the traps in
Bose-Einstein condensates in optical lattices \cite{ng2008}. Possibly
also other systems with large coherence times, like optical waveguide
lattices \cite{hagai2008}, could be used to study this effect.

\section{Acknowledgments}

OM thanks Jens Harting for valuable support in computational matters. We
gratefully acknowledge support from the Deutsche
For\-schungs\-ge\-mein\-schaft (DFG) and the Fonds der Chemischen
Industrie.

\bibliographystyle{eplbib}

\end{document}